\documentclass[english,aps,twocolumn, floatfix,showpacs,pra,preview]{revtex4}
\usepackage[T1]{fontenc}
\usepackage[utf8]{inputenc}
\usepackage{units}
\usepackage{mathtools}
\usepackage{amsmath}
\usepackage{amssymb}
\usepackage{cancel}
\usepackage{stackrel}
\usepackage{subfig}
\usepackage{graphicx}
\usepackage{color,soulutf8, ulem}
\usepackage{hyperref}

\makeatletter
\newcommand{\ket}[1]{|#1\rangle}
\newcommand{\bra}[1]{\langle#1|}

\newcommand{\be}{\begin{equation}}
\newcommand{\ee}{\end{equation}}

\newcommand{\sza}{\hat{S}_z^{(a)}}
\newcommand{\szb}{\hat{S}_z^{(b)}}

\newcommand{\bb}[1]{\left( #1\right)}

\makeatother

\usepackage{babel}
\begin{document}

\title{Evolution of entanglement under an Ising\textendash like Hamiltonian  with particle losses}


\author{Konrad Szyma\'{n}ski}
\affiliation{Center for Theoretical Physics PAS, Aleja Lotnik\'ow 32/46, 02-668 Warszawa, Poland\\
Marian Smoluchowski Institute of Physics, ul. Łojasiewicza 11, 30-348 Kraków, Poland}

\author{Krzysztof Paw\l owski}
\affiliation{Center for Theoretical Physics PAS, Aleja Lotnik\'ow 32/46, 02-668 Warszawa, Poland}
\date{\today}

\begin{abstract}
We present analytical compact solution for the density matrix and all correlation functions of two collective-macroscopic spins evolving via Ising-like Hamiltonian in the presence of particle losses.
The losses introduce non-local phase noise which destroys highly entangled states arising in the evolution. On the other hand, the states appearing at relatively short timescales, possessing EPR-like entanglement will survive. 
Applying our solutions to the recently proposed scheme to entangle two Bose-Einstein condensates, we estimate the optimal number of atoms for EPR correlations.
\end{abstract}

\maketitle
\section{Introduction}
There exist a number of experiments in which entanglement between massive particles is generated. Many setups are on the list: ultracold gases \cite{esteve2008, Riedel2010, Gross2010}, atoms trapped in the resonant cavities \cite{Haas:2014kb, SchleierSmith:2010eo, schleier-smith:2010vn} or superconducting qubits \cite{Vlastakis607}. 
The entanglement in each of these experiments stems from different physical mechanisms. Ultracold atoms entangle simply  due to atom-atom collisions \cite{Sorensen2001,esteve2008, Riedel2010, Gross2010}. 
The diluted thermal atoms were made to interact by a common mode of light in resonant cavities \cite{Haas:2014kb, SchleierSmith:2010eo, schleier-smith:2010vn}. 

Although the setups were very different, in all cases the emergence of entanglement can be understood, at least qualitatively, within a common simple theoretical model, so called one-axis twisting scheme \cite{Kitagawa1993}. 
In this scheme the state of the system is expressed as a macroscopic collective spin $\bm{S}$ of the length proportional to the number of particles $N$. The quantum correlations result from an unitary evolution due to the  Hamiltonian proportional to $ \hat{S}_z^2$, where $\hat{S}_z$ is the $z$-th component of the collective spin.
One of the mentioned above experiments has been extended \cite{Wang1087}, and other are proposed to be extended
\cite{Opanchuk2012, hadrien2013, Rosseau2014}, to  a setup described by two effective collective spins $S^a$ and $S^b$, which due to nonlocal Hamiltonian $S^a_z S^b_z$ would evolve into non-local entangled states.
If one manage to prepare initially both collective spins to be along $x$ axis, then due to the latter non-local Hamiltonian they would evolve first to state possessing EPR-type of entanglement and then to  more exotic non-local macroscopic superpositions, as so called "Schr\"odinger cat in two boxes" \cite{hadrien2013, byrnes2013devil, Wang1087}.

Preparation of such state is of fundamental interest, as it shows at macroscopic level the early quantum mechanics paradoxes \cite{epr}.
The scheme is also considered from the quantum-computation perspective \cite{pyrkov2013}.

As one would like to extend the existing experiments into this new non-local regime, it is critical to ask again for robustness of the scheme in the presence of decoherence.
In this manuscript we will investigate the role of collective particle losses, process typically found in Bose-Einstein condensates.
The system we discuss, with one-body losses included via standard master equation, has a favorable theoretical feature --  it has a simple compact analytical solutions, for all correlation functions and for all terms of the density matrix of the system.

After presenting our model of the non-local evolution with particle losses in Sec. \ref{sec:model} we will discuss a method  of generating functions with which we find the analytical solutions, Sec. \ref{sec:methods}.
In Sec. \ref{sec:entanglement} we show the evolution of  the linear entropy and so called EPR condition \cite{Reid2009, hadrien2013, Cavalcanti2009, Cavalcanti2011}, known in the Quantum Information community under the name "steering condition".
In Sec. \ref{sec:quantum-trajectories} we use the quantum trajectory method to gain an insight into the lossy dynamics.
It is shown that the losses result in the non-local phase noise - loosing an atom in system "a" leads through the non-local evolution to a noise in the system "b".
The entanglement captured by EPR condition is less sensitive to the  action of losses, what is qualitatively understood by discussing the phase noise introduced in the frame of the quantum trajectories, given in Sec. \ref{sec:quantum-trajectories}.
We focus on this type of entanglement in the last section, where we include phenomenologically two- and three-body losses to estimate the optimal conditions for an experiment in ultracold gases.

\section{Model \label{sec:model}}
We consider nonlinear evolution of two groups (denoted with indices "a" and "b")
of (pseudo)spins $1/2$. We will assume, that in the initial state
each from $N+M$ spins is in the state $\frac{\ket{0}+\ket{1}}{\sqrt{2}}$,
namely 
	\begin{equation}
	\ket{\psi(0)}=\bigotimes_{i=1}^{N}\frac{\ket{0}_{a}+\ket{1}_{a}}{\sqrt{2}}\bigotimes_{i=1}^{M}\frac{\ket{0}_{b}+\ket{1}_{b}}{\sqrt{2}}.
	\label{eqn:psi0}
	\end{equation}
The Hamiltonian under consideration is
	\begin{equation}
	H=\chi(S_{z}^{a})^{2}+\chi(S_{z}^{b})^{2}-\chi_{ab}S_{z}^{a}S_{z}^{b},
	\label{eqn:ham0}
	\end{equation}
where $S_{z}^{a}=\frac{1}{2}\sum_{i=1}^{N}\sigma_{iz}^{a}$ is sum
of $z$-Pauli matrices of all of $N$ spins in the group "a".
Similar definition holds for $S_{z}^{b}$, but with $M$ spins. 
In addition to the unitary evolution we consider dissipation. This
is an important point, as in general entanglement existing between
two subsystems is quite susceptible to very small deviations from
unitarity (or in other words, to a weak entanglement with an environment). Due
to the resulting decoherence, the state of the system will be given
by a density operator, which we assume to obey master equation
	\begin{equation}
	\frac{d\hat{\rho}}{dt}=i[\hat{\rho},H]+L[\hat{\rho}].\label{eq:master}
	\end{equation}
As the source of the dissipation we will assume one body losses modeled
by the following Lindblad superoperator

\be
L[\hat{\rho}]=\sum_{\epsilon}\Gamma_{\epsilon}\left(\hat{a}_{\epsilon}\hat{\rho}\hat{a}_{\epsilon}^{\dagger}+\hat{b}_{\epsilon}\hat{\rho}\hat{b}_{\epsilon}^{\dagger}-\frac{1}{2}\left\{ \hat{\rho},\hat{a}_{\epsilon}^{\dagger}\hat{a}_{\epsilon}+\hat{b}_{\epsilon}^{\dagger}\hat{b}_{\epsilon}\right\} \right),
\label{eq:lindblad}
\ee
where $\hat{a}_{\epsilon}$ ($\hat{b}_{\epsilon}$) is a bosonic operator which anihilates a single
atom in state $\epsilon=0,1$ from mode "a" ("b"). This model is particularly adequate to Bose-Einstein
condensates, for which the term \eqref{eq:lindblad}
describes losses due to interaction with external particles, coming from 
residue air in the vacuum chamber.

As the initial state belongs to the symmetric subspace and dissipation
is expressed with collective operator only, hence the appropriate
basis will be the Fock basis: 
\be
\ket{n_{0},n_{1},m_{0},m_{1}}=\frac{\left(\hat{a}_{0}^{\dagger}\right)^{n_{0}}\left(\hat{a}_{1}^{\dagger}\right)^{n_{1}}\left(\hat{b}_{0}^{\dagger}\right)^{m_{0}}\left(\hat{b}_{1}^{\dagger}\right)^{m_{1}}}{\sqrt{n_{0}!n_{1}!m_{0}!m_{1}!}}\ket{\cancel{0}},
\label{eq:Fock}
\ee
where $\ket{\cancel{0}}$ is the vacuum.

The density matrix written in this basis reads:
\begin{equation}
\hat{\rho}= \sum_{\substack{n_0n_1m_0m_1\\k_0 k_1 l_0 l_1}} \rho_{k_0k_1l_0l_1}^{n_0n_1m_0m_1} \ket{n_{0}n_{1}m_{0}m_{1}}\bra{k_{0}k_{1}l_{0}l_{1}},
\label{eq:density-matrix-terms}
\end{equation}
where the ranges of the indices within the sum are for kets, "a": $n_0=0,\ldots,N $, $n_1=0,\ldots,N-n_0 $, for kets "b"   $m_0=0,\ldots, M$, $m_1=M-m_0$ and  the indices for bras  are constrained analogously.

Another states which will be helpful in the next sections are the so called phase states \cite{Leggett1991, alice1998}, here defined as:
	\begin{eqnarray}
	\ket{\phi_a\phi_b} &\equiv& \bigotimes_{i=1}^{N}\frac{e^{-i\phi_a/2}\ket{0}_{ia}+e^{i\phi_a/2}\ket{1}_{ia}}{\sqrt{2}} \nonumber\\
	 & &\bigotimes_{j=1}^{M}\frac{e^{-i\phi_a/2}\ket{0}_{jb}+e^{i\phi_a/2}\ket{1}_{jb}}{\sqrt{2}}.
	 \label{eq:phase-state}
	\end{eqnarray}

\section{Methods\label{sec:methods}}
The master equation \eqref{eq:master} written in the Fock basis leads to a problem which complexity grows with the number of particles: a set of numerous ($N^3M^3$) coupled ordinary differential equations on the density matrix terms $\rho_{k_0k_1l_0l_1}^{n_0n_1m_0m_1}$ needs to be solved.
There exists however a mathematical technique well suited for this class of problems:
method of characteristic functions. It has been already applied to
problems involving Bose-Einstein condensates \cite{pawlowskiBackground}, our
approach is a natural extension of the previous research. We will
start with a simple example explaining the basic concepts of the method,
which we will later apply to our system.
\subsection{Simple example\label{sec:simple-example}}
In this section, we will solve a simple set of differential equations:
\begin{equation}
\frac{dp_{n}}{dt}=\Gamma\left((n+1)p_{n+1}-np_{n}\right),
\label{eq:example1}
\end{equation}
where $\{p_n\}$ are the unknown function of the time and $n=0, 1, \ldots, \infty$.
There are numerous approaches to this (elementary) problem. One which
renders it extremely simple is consideration of the following polynomial
of $x$: 
\[
P(x,t)=\sum_{n}x^{n}p_{n}(t).
\]
We can reconstruct time evolution of this expression by multiplying
each of equations \eqref{eq:example1} by $x^{n}$ and summing over $n$:
\[
\sum_{n}x^{n}\frac{dp_{n}(t)}{dt}=\Gamma\sum_{n}x^{n}\left((n+1)p_{n+1}(t)-np_{n}(t)\right),
\]
which can be rewritten in the form 
\[
\frac{\partial P(x,t)}{\partial t}=\Gamma\left(1-x\right)\frac{\partial P(x,t)}{\partial x}.
\]
Hence instead of set of equations we have arrived to just one first
order partial differential equation (PDE).

Solution of this particular equation is easy to guess, but to solve our master equation \eqref{eq:master} we will deal with
the more complicated ones, which will be solved with the method of characteristics.

Having polynomial $P(x,t)$ we can reconstruct the original $p_{n}(t)$:
from the very definition of $P(x,t)$, where $p_{n}$ is just the coefficient
at $x^{n}$ and: 
\[
p_{n}(t)=\frac{1}{n!}\left.\frac{\partial^{n}P\left(x,t\right)}{\partial x^{n}}\right|_{x=0}.
\]

\subsection{Application of the method in our system}

Writing out the master equation \eqref{eq:master}  yields a rather
long expression even with the usage of Fock basis. Fortunately it is possible to
 substantially simplify the problem.

By explicit calculation one can find that  the master equation \eqref{eq:master}
couples only the density matrix terms $ \rho_{k_0k_1l_0l_1}^{n_0n_1m_0m_1}$
for which the indices differences:
	\begin{align*}
	\delta_{a} & =\left(n_{0}+n_{1}\right)-\left(k_{0}+k_{1}\right),\\
	\delta_{b} & =\left(m_{0}+m_{1}\right)-\left(l_{0}+l_{1}\right).
	\end{align*}
are equal. There is a reason to that:
Physically, as the particles in our system are massive, there are
no coherences between different total particle number states and
thus, $\delta_{a}+\delta_{b}=0$. On top of that, we start the evolution
with a product state, $\hat{\rho}(t=0)=\hat{\rho}_{a}\otimes\hat{\rho}_{b},$
so such coherences are absent in both subsystems in
fact, $\delta_{a}=\delta_{b}=0$. As the evolution preserves $\delta_{a},\delta_{b},$
we may write
\begin{align*}
(n_{0},n_{1},m_{0},m_{1}) & =(x,y+z,u,v+r),\\
(k_{0},k_{1},l_{0},l_{1}) & =(x+z,y,u+r,v).
\end{align*}
In this way, instead of $8$ indices used to indicate matrix elements,
we will use only $6$ indices. The master equation couples only these
matrix elements which has the same indices $z$ and $r$ being offsets
from diagonal.

From now on, we will use this sub-basis. 
The derivation of the characteristic function equation is straightforward but lengthy, hence we will just write down the
final results (for a more detailed version please refer to the appendix).
We introduce a family of characteristic functions
\begin{eqnarray}
h^{z,r}(X,Y,U,V,t)=\quad\sum_{x,y,u,v}X^{x}Y^{y}U^{u}V^{v}\quad\quad~\nonumber& \\\sqrt{\frac{(x+z)!(y+z)!(u+r)!(v+r)!}{x!y!u!v!}}\rho_{x+z,y,u+r,v}^{x,y+z,u,v+r}&,
\label{eq:def-gener_fct}
\end{eqnarray}
indexed by pair $(z,r)$ - the offset from diagonal elements
used in defining sum. After laborious derivation final equation for
$h^{z,r}$ is
\begin{align}
\frac{\partial h^{z,r}}{\partial t}= & \left(-\beta^{(r,z)}_0X+\Gamma_{0}\right)\frac{\partial h^{z,r}}{\partial X}+
\left(\beta^{(r,z)}_1 Y+\Gamma_{1}\right)\frac{\partial h^{z,r}}{\partial Y}+ \nonumber\\
&  \left(-\beta^{(z,r)}_0U+\Gamma_{0}\right)\frac{\partial h^{z,r}}{\partial U}+
\left(\beta^{(z,r)}_1 V+\Gamma_{1}\right)\frac{\partial h^{z,r}}{\partial V} \nonumber\\
& -  2\left(\Gamma_{0}+\Gamma_{1}\right)(z+r)h^{z,r}. \label{eq:characteristic-function}
\end{align}

where 
\begin{equation}
\beta^{(p,q)}_{\epsilon}=2(i (p \chi_{ab} + 2 q \chi) \pm \Gamma_\epsilon),
\end{equation}
in which "$+$" sign corresponds to $\epsilon=0$, and "$-$" to $\epsilon=1$.


The initial state
	\[
	\ket{\psi(0)}=2^{-\frac{M+N}{2}}\sum_{n=1} \sqrt{\binom{N}{n}}\sqrt{\binom{M}{m}}\ket{n, N-n,m,M-m}.
	\]
leads to the following initial condition for the characteristic function $h^{z,r}(X,Y,U,V,0)$:
	\begin{equation}
	h^{z,r}|_{t=0}=\frac{2^{-(M+N)}M!N!}{(M-r)!(N-z)!}\left(X+Y\right)^{N-z}\left(U+V\right)^{M-r}.
	\label{eq:initialc-characteristic-function}
	\end{equation}
We solve the partial differential equation \eqref{eq:characteristic-function} with the initial condition \eqref{eq:initialc-characteristic-function} using the method of characteristics. As in the simple example 
we find a path in the space of parameters, here $X,Y,U,V, t$, on which the $h^{z,r}$ changes exponentially \footnote{
We will omit this part \textemdash{}
the calculation, while simple, is lengthy and generally the easiest
method is to employ Mathematica to perform this laborous task.}.
As described in the next section the characteristic functions are particularly well suited to evaluate averages of the polynomial of the spin component function, making the method a powerful tool.

\subsection{Quantum averages}
In order to evaluate average of some operator, one can express the average as a function of the density matrix terms.
In principle it is possible to calculate the matrix elements from the
characteristic function,
\[
\rho_{x+z,y,u+r,v}^{x,y+z,u,v+r} \propto
 \left.\frac{\partial^{x+y+u+v}h^{z,r}}{\partial X^{x}\partial Y^{y}\partial U^{u}\partial V^{v}}\right|_{X=Y=U=V=0},
\]
The straightforward calculation is however inefficient (there are $O(N^{3}M^{3})$ matrix elements in general,
but not all of them are needed) and often not necessary: pseudospin-related averages can be extracted directly from
the characteristic function. Let us take, for instance, operator $S_{x}^{a}=\frac{a_{1}^{\dagger}a_{0}+a_{0}^{\dagger}a_{1}}{2}$:
its average value is $\textrm{Tr}\left\{ \hat{\rho}S_{x}^{a}\right\} $.
The $\textrm{Tr}\left\{ \hat{\rho}a_{1}^{\dagger}a_{0}\right\} $
term is equal to $\sum_{x,y,u,v}\sqrt{y(x+1)}\rho_{x-1,y,u,v}^{x,y-1,u,v}$, which can be easily obtained from one of the characteristic functions:
\begin{equation}
\left\langle a_{1}^{\dagger}a_{0}\right\rangle =\textrm{Tr} \left\{\hat{\rho}a_{1}^{\dagger}a_{0}\right\}=\left. h^{-1,0}\right|_{X=Y=U=V=1}.
\end{equation}
All important correlation function we are calculated follow the same path.
We give explicit form of chosen quantum averages in the Appendix \ref{app:quantum-averages}.
\section{Entanglement in the system \label{sec:entanglement}}
\subsection*{Linear entropy}

\begin{figure}
\includegraphics[width=0.4\textwidth]{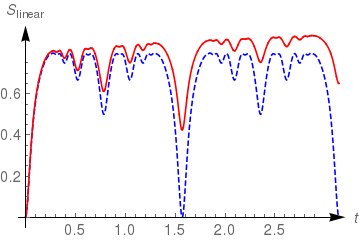}
\includegraphics[width=0.4\textwidth]{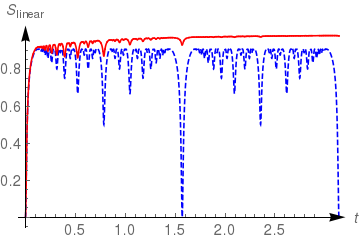}
\caption{
Linear entropy as a function of time in case: without losses $\Gamma_{1}=\Gamma_{0}=0.0$ (red solid line),  with   $\Gamma_{1}=\Gamma_{0}=0.01$ (blue dashed line).
Top: $N=10$, bottom $N=50$. Other parameters $\chi_{ab}=\chi$ and $N=M=10$.
\label{fig:s1}}
\end{figure}

In case without losses one can quantify the entanglement using the linear entropy defined as
\begin{equation}
S_{\rm lin}=1-\operatorname{Tr} \sigma^2=1-\sum_{x,y,z}\left|\sigma_{x+z,y}^{x,y+z}\right|^{2},
\label{eq:linear-entropy}
\end{equation}
where $\sigma = \operatorname{Tr}_b\left\{ \rho  \right\}$ is the reduced density matrix.

We have not found a simple equation for the linear entropy as a function
of $h^{z,r}$ functions. Instead, it is possible
to compute all the matrix elements of a partially traced matrix:
\[
\sigma_{x+z,y}^{x,y+z}=\sqrt{\frac{1}{x!y!(x+z)!(y+z)!}}\left.\frac{\partial^{x+y}h^{z,0}}{\partial X^{x}\partial Y^{y}}\right|_{X=Y=0,U=V=1}.
\]
With the help of these reduced matrix elements, we evaluate the linear entropy for small $N, M$ (of order 10) using directly the definition \eqref{eq:linear-entropy}.

We illustrate evolution of this quantity in Fig. \ref{fig:s1}. In
case without losses, $\Gamma_{0}=\Gamma_{1}=0$ this entropy, similarly
to von Neuman entropy \cite{hadrien2013, byrnes2013devil} and negativity \cite{Rosseau2014} has a fractal-like
structure, independent on the value of $\chi$ (see Appendix in \cite{hadrien2013}). This is reminiscent of the so called devil's staircase known from the Ising model \cite{devilsStaircase}. 
Each peak indicates a superposition of a few phase states \eqref{eq:phase-state}, with the cat-like state appearing at $\chi_{ab}t=\pi$.
Precisely at $t=\pi/\chi_{ab}$, the state is a superposition of 
$\ket{0,0}_{\text{ph}}$,
$\ket{0,\pi}_{\text{ph}}$,
$\ket{\pi,0}_{\text{ph}}$,
$\ket{\pi,\pi}_{\text{ph}}$
(the eigenstates of $S_x^aS_x^b$ operator) with amplitudes depending on $\chi$.
As an example, in the case of $\chi=0$ the state reads
	\begin{equation}
	    \ket{\Psi(t)}=\frac{\ket{00}_{\text{ph}}+\ket{0\pi}_{\text{ph}}+\ket{\pi 0}_{\text{ph}}-\ket{\pi\pi}_{\text{ph}}}{2}.
	    \label{eq:sch-cat-in-two-boxes}
	\end{equation}
Although the linear entropy is no a direct quantifier of entanglement when the losses are present, still its dramatic change  after one (on average) loss event, as shown in Fig. \ref{fig:s1}, reveals that the entangled states at long evolution times ($\chi_{ab} t \gg 1/\sqrt{N}$) are, as expected, vulnerable to decoherence. Hence in the next sections we focus on short evolution times.

\subsection{Short times: EPR like entanglement}
\begin{figure}
    \includegraphics[width=0.45\textwidth]{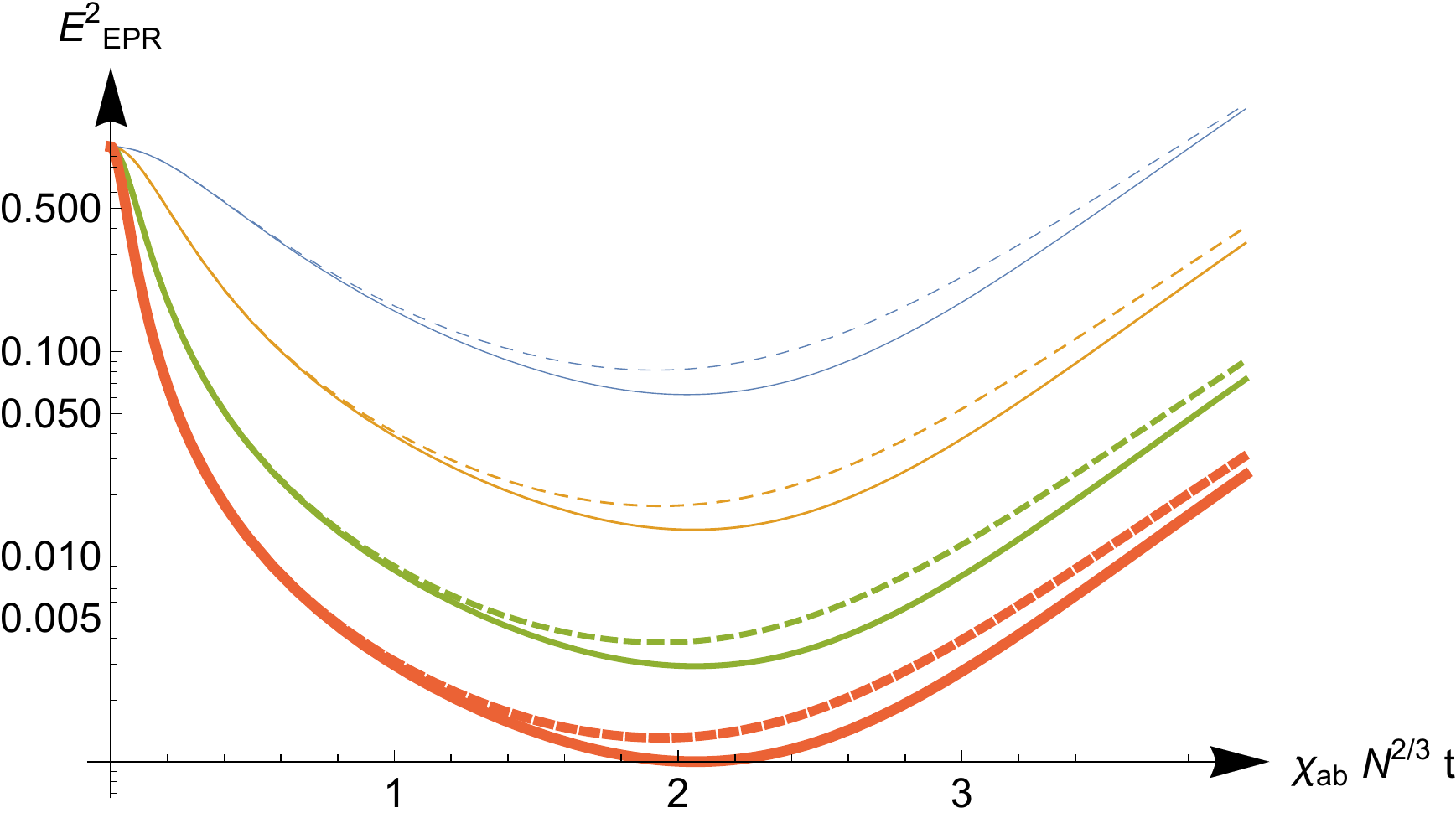}
    \caption{ The entanglement criterion $E_{\rm EPR}^2$ for various number of particles as a function of rescaled time.
        For comparison, $E^2_{\rm EPR}$ calculated disregarding losses is imposed as dashed lines.
The thickness of line refers to $N$: from least to most thick, $N=10^2,10^3,10^4,5\times10^4$.
Other parameters: $\Gamma_{1}=\Gamma_{0}=\chi_{ab}$ and $\chi=0$. The optimal angles are $\alpha=\beta=0$.
\label{fig:e1}}
\end{figure}

We employ entanglement criterion devised in \cite{wiseman2007, Opanchuk2012, Cavalcanti2011} which
is constructed from simple observable quantities and it is sensitive
to EPR-like correlations \cite{hadrien2013}.

The entanglement criterion is rooted in the following reasoning. Let
us consider two observables, $\hat{X}$ and $\hat{P}$. Then for a single
isolated system, the expression $L=\left\langle \left(\hat{X}-A\right)^{2}\right\rangle \left\langle \left(\hat{P}-B\right)^{2}\right\rangle $
is minimized by a choice $A=\left\langle \hat{X}\right\rangle $ and
$B=\left\langle \hat{P}\right\rangle $. The Heisenberg inequality
tells us, that even for this optimal choice of the numbers $A$ and
$B$, the expression $L$ has to be larger than $\frac{1}{4}\left|\left\langle \left[\hat{X},\hat{P}\right]\right\rangle \right|^{2}$.

The situation changes if nonclassical correlations i.e., entanglement
between system "a" and another system
"b" are allowed. As in the seminal paper
of Einstein Podolsky and Rosen \cite{epr}, there exist states in which
after every measurement performed in "b"
we know precisely the current value of either position or momentum
in "a". It is then possible to construct
a better estimates for $\hat{X}$ and $\hat{P}$, based on observations
made in the second subsystem. Mathematically, one can violate
the inequality 
\begin{equation}
\left\langle \left(\hat{X}^{a}-\hat{X}_{\rm inf}^{a}\right)^{2}\right\rangle \left\langle \left(\hat{P}^{a}-\hat{P}_{\rm inf}^{a}\right)^{2}\right\rangle \ge\frac{1}{4}\left|\left\langle \left[\hat{X}^{a},\hat{P}^{a}\right]\right\rangle \right|^{2}\label{eq:ineq}
\end{equation}
where the new estimators $\hat{X}_{\rm inf}^{a}$, $\hat{P}_{\rm inf}^{a}$ are
functions of observables in the second subsystem (in our case linear
dependence is enough to prove the existence of entanglement): 
\begin{align*}
\hat{X}_{\rm inf}^{a} & =q^{a}+r^{a}\hat{X}^{b},\\
\hat{P}_{\rm inf}^{a} & =p^{a}+s^{a}\hat{P}^{b}.
\end{align*}
We can define the parameter 
\begin{equation}
E_{EPR}^{2}=\frac{4\left\langle \left(\hat{X}^{a}-\hat{X}_{inf}^{a}\right)^{2}\right\rangle \left\langle \left(\hat{P}^{a}-\hat{P}_{inf}^{a}\right)^{2}\right\rangle }{\left|\left\langle \left[\hat{X}^{a},\hat{P}^{a}\right]\right\rangle \right|^{2}},
\label{eq:def-eepr}
\end{equation}
which, if less than $1$, serves as an indicator of EPR entanglement.
The optimal parameters $q^{a},r^{a},p^{a},s^{a}$, minimizing the $E_{EPR}^{2}$ are defined by
\[
\begin{array}{cccc}
q^{a}=\left\langle \hat{X}_{a}\right\rangle -\frac{\textrm{covar}(\hat{X}^{a},\hat{X}^{b})}{\left\langle \left(\hat{X}^{b}-\left\langle \hat{X}^{b}\right\rangle \right)^{2}\right\rangle }\left\langle \hat{X}_{b}\right\rangle  & , & r^{a}= & \frac{\textrm{covar}(\hat{X}^{a},\hat{X}^{b})}{\left\langle \left(\hat{X}^{b}-\left\langle \hat{X}^{b}\right\rangle \right)^{2}\right\rangle },\\
p^{a}=\left\langle \hat{P}_{a}\right\rangle -\frac{\textrm{covar}(\hat{P}^{a},\hat{P}^{b})}{\left\langle \left(\hat{P}^{b}-\left\langle \hat{P}^{b}\right\rangle \right)^{2}\right\rangle }\left\langle \hat{P}_{b}\right\rangle  & , & s^{a}= & \frac{\textrm{covar}(\hat{P}^{a},\hat{P}^{b})}{\left\langle \left(\hat{P}^{b}-\left\langle \hat{P}^{b}\right\rangle \right)^{2}\right\rangle },
\end{array}
\]
where 
\[
\textrm{covar}(\hat{A},\hat{B})=\left\langle \hat{A}\hat{B}\right\rangle -\left\langle \hat{A}\right\rangle \left\langle \hat{B}\right\rangle .
\]
In our case the simplest observables are pseudospins 
	\begin{equation}
	\begin{array}{ccccc}
	\hat{X}^{a}= & S_{\alpha}^{a}&, \quad &\hat{P}_{\alpha}^{a}= & S_{\alpha+\pi/2}^{a},\\
	\hat{X}^{b}= & S_{\beta}^{b}&,\quad & \hat{P}_{\alpha}^{b}= & S_{\beta+\pi/2}^{b},
	\end{array}
	\label{eq:x-to-spin}
	\end{equation}
were we used the notation
	\begin{align}
	S_{\alpha}^{i} & =\cos\alpha S_{x}^{i}+\sin\alpha S_{y}^{i},\\
	S_{\alpha+\pi/2}^{i} & =\sin\alpha S_{x}^{i}-\cos\alpha S_{y}^{i}.
	\label{eq:def_Salpha}
	\end{align}
To find the minimal value of $E_{\rm EPR}^2$ one should plug the form \eqref{eq:def_Salpha} into \eqref{eq:def-eepr} and then minimize with respect to the free parameters $\alpha$ and $\beta$, used in the definition \eqref{eq:x-to-spin}.

With the method of generating function we have obtained exact analytical
formulas for all average fluctuations and covariances of the spin components, also in the case with particle losses.
In Fig. \ref{fig:e1} we show the resulting $E_{EPR}^{2}$ as a function of time for different number of atoms.
As we show, even for substantial losses ($\Gamma$s comparable with other parameters , which translates to $\operatorname{const}\times N^{1/3}$ particles lost by the time of $E_{EPR}$ minimum) the entanglement is preserved.
This is due to timescales at which this entanglement appears, as intuitively explained at the end of the next section.

\section{Quantum trajectories \label{sec:quantum-trajectories}}
One can get a better understanding on what is happening during the evolution using  quantum trajectories
method \cite{belavkin1990, dalibard1992}, sometimes interpreted as a way to generate a  collection of single experimental realizations.

The basic notions in quantum trajectories method are
the effective Hamiltonian $H_{\mathrm{eff}}$ and nonunitary transition operators $\left\{ A_{i}\right\} $ ( often called quantum jump operators),
an instantaneous quantum state of a single realization (a vector in pertinent
Hilbert space) and its trajectory over time. Basic evolution is simple:
for a state $\ket{\psi(t)}$, state in the next moment of time $\ket{\psi(t+\Delta t)}$
can be calculated in a probabilistic manner: 
\begin{itemize}
\item a transition associated with operator $A_{i}$ will happen with probability
$p_{i}=\Delta t\bra{\psi(t)}A_{i}^{\dagger}A_{i}\ket{\psi(t)}$,
after which the state reads \\
	\[
	\ket{\psi(t+\Delta t)}:=\frac{A_{i}\ket{\psi(t)}}{\left|\bra{\psi(t)}A_{i}^{\dagger}A_{i}\ket{\psi(t)}\right|},
	\]
\item Hamiltonian-like evolution will happen with probability $1-\sum p_{i}$,
after which the state reads
	\begin{equation}
	\ket{\psi(t+\Delta t)}:=\frac{\exp\left(-\Delta t\left(\sum A_{i}^{\dagger}A_{i}/2+iH\right)\right)\ket{\psi(t)}}{\left|\exp\left(-\Delta t\left(-\sum A_{i}^{\dagger}A_{i}/2+iH\right)\right)\ket{\psi(t)}\right|},
	\label{eq:eff-evolution}.
	\end{equation}
	where
	\begin{equation}
	H_\text{eff}=\sum A_{i}^{\dagger}A_{i}/2i+H
	\end{equation}
	is the effective Hamiltonian of the system. Note that this object is not Hermitian unless $\sum A_{i}^{\dagger}A_{i}=0$.
\end{itemize}

Please note that even if we are sure that transition did not occur (the case when Eq. \eqref{eq:eff-evolution} is applied),
it has an effect on the system: effectively, the contribution to the
state vector from the basis vectors undergoing transition is reduced.
In experiment, this corresponds to post-selection of runs in which
transition did not happen. Naturally, statistics of the final states
is modified as compared to case without losses and this is precisely
the effect seen here.

The connection with usual treatment using density matrices is straightforward:
density matrix of the system is recovered after averaging the state
projectors, $\ket{\psi(t)}\bra{\psi(t)}$ over multiple realizations.
This approach could in principle be used to calculate the density
matrix in an efficient manner (the number of differential equations
scales linearly with the dimensionality of the Hilbert space, which
corresponds to linear speedup compared to number of density matrix
elements), but its power lies elsewhere: within the quantum trajectories
framework, it is possible to calculate what is the state of the system
if we know, for instance, that one particular transition $A_{i}$
occurred. 

In this paper, we will employ the quantum trajectories method to show the "microscopic" mechanisms, caused by losses and leading to decoherence as it was done in case of a single bimodal BEC \cite{pawlowski2013}.
The jump operators associated with the master equation \eqref{eq:master}  are: $A_0^{(1)} = \sqrt{\Gamma_0} \hat{a}_0$ and $A_1^{(1)} = \sqrt{\Gamma_1} \hat{a}_1$ for the subsystem "a" and $A_2^{(1)} = \sqrt{\Gamma_0} \hat{b}_0$ and $A_3^{(1)} = \sqrt{\Gamma_1} \hat{b}_1$ for "b".
It is enough to show the effect of losses in the subspace with $N+M-1$ atoms, i.e. after loss of one particle.
The (unnormalized) single trajectory at time $t$ but with particle lost at time $t_1<t$ due to the $i$--th jump operator reads 
\begin{equation}
\begin{split}
\ket{\psi(t | i, t_1)} =& e^{ \left[- i H - \frac{1}{2} \sum_i \bb{A_i^{(1)}}^{\dagger} A_i^{(1)} )\right] (t-t_1)} A_i^{(1)}\\
  & e^{ \left[- i H - \frac{1}{2} \sum_i \bb{A_i^{(1)}}^{\dagger} A_i^{(1)} \right] t_1} \ket{\psi(0)}.
\end{split}
\end{equation}
Omission of normalization is beneficial in the next point, in which the part of density matrix at $t$ with a single lost particle is calculated. This part is exactly an average over the times at which this particle has been lost with equal weights:
\begin{equation}
\label{eqn:rhoprim}
\rho'(t) = \int_{0}^{t}\ket{\psi(t |i, t_1)}\bra{\psi(t |i, t_1)} \textrm{d}t_{1},
\end{equation}
as the length of unnormalized $\ket{\psi(t |i, t_1)}$ squared is proportional to the probability of quantum jump occurring at $t_1$.

The trajectory $\ket{\psi(t |i, t_1)}$ can be written as
\begin{equation}
\ket{\psi(t |i, t_1)}\propto \exp\left( \pm i t_1 \left( \frac{\chi_{ab}}{2} S^b_z - 3\chi S^a_z \right) \right) \bb{A_i^{(1)} \ket{\psi(t)}},
\label{eq:psitt1}
\end{equation}
where the "+" sign corresponds to $i=0$ and "--" to $i=1$.
In Eq. \eqref{eq:psitt1}  the proportionality factor does not depend on $t_1$ and $\ket{\psi(t)}$ is the "original" state vector, as if the evolution remained unperturbed by quantum jumps. 
 The density matrix in subspace with one lost particle can be obtained according to Eq. \eqref{eqn:rhoprim}. 

The prefactor $\exp\left(i t_1 \left( \frac{\chi_{ab}}{2} S^b_z - 3\chi S^a_z \right) \right) $ corresponds to nonlocal, $t_1$--dependent rotation of the state vector, which after integration destroys the state coherence. 
This is probably the major mechanism underlying the decoherence for short times, already known  under the name "phase noise". In every quantum trajectory the evolving state is "entangled" in a similar degree. The only effect of losses is the rotation over an angle, which depends on the random time $t_1$ at which the atom has been lost. This is the averaging over the random time $t_1$ which results in the deterioration of the entanglement. 
This effect can be seen by calculating the cut of the Husimi function $Q$:

	\[
	Q(\phi_a,\phi_b)=\bra{\phi_a,\phi_b}\rho'\ket{\phi_a,\phi_b},
	\]
with respect to the phase states \eqref{eq:phase-state}, i.e. equatorial product states, eigenvectors of $S_{\phi_a}^{a}S_{\phi_b}^{b}$:
	\[
	\ket{\phi_a,\phi_b}=e^{i \phi_a\sza} e^{i \phi_b\szb}\sum\sqrt{\binom{N}{n}\binom{M}{m}}\ket{n,N-n,m,M-m}.
	\]

\begin{figure}
\subfloat[$Q(\alpha,\beta)$ in the subspace where no particle is lost.]{\includegraphics[width=0.15\textwidth]{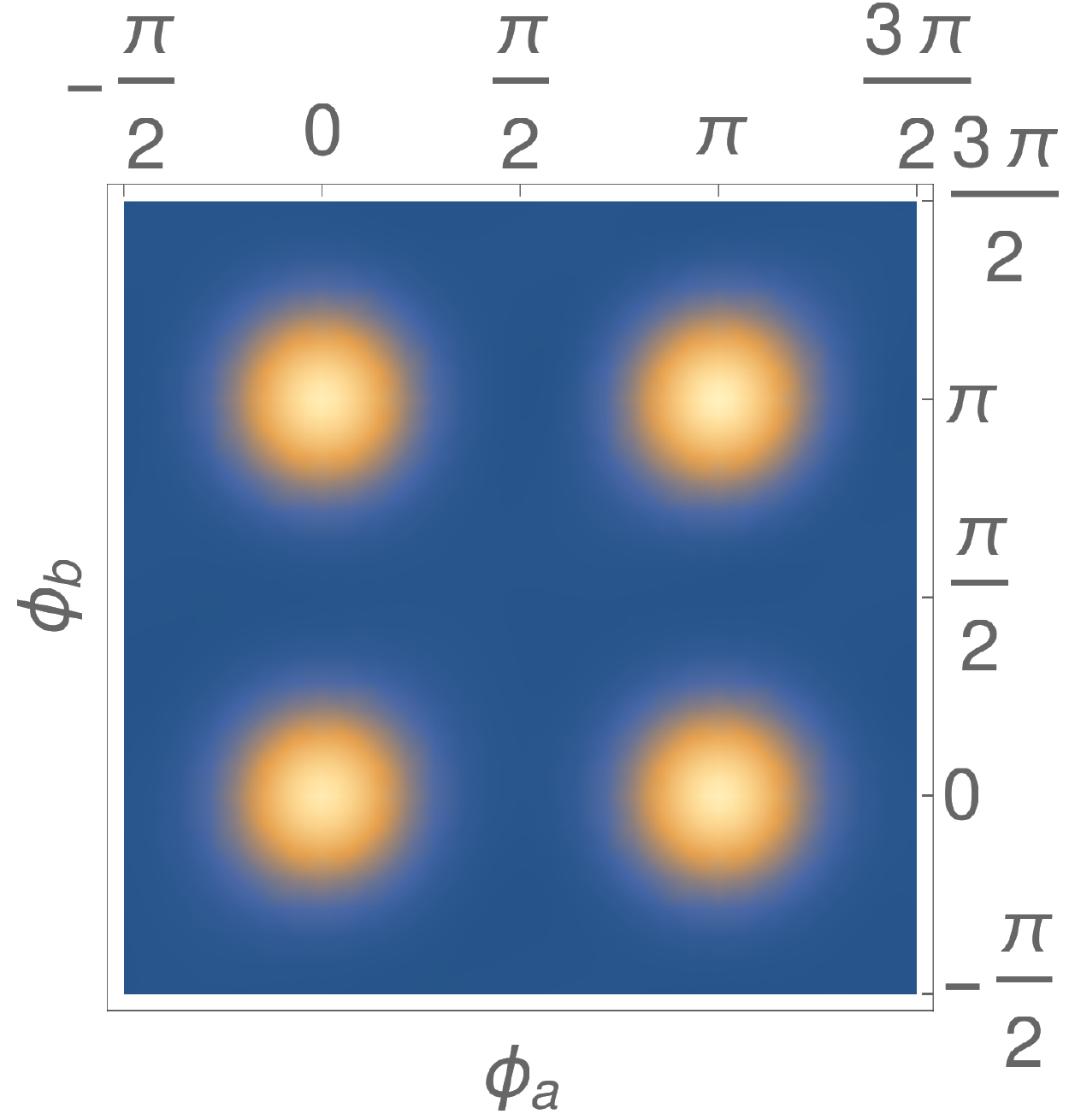}

}
\subfloat[$Q'(\alpha,\beta)$ in the subspace where one particle in the mode 0 of condensate "a" is lost.]{\includegraphics[width=0.15\textwidth]{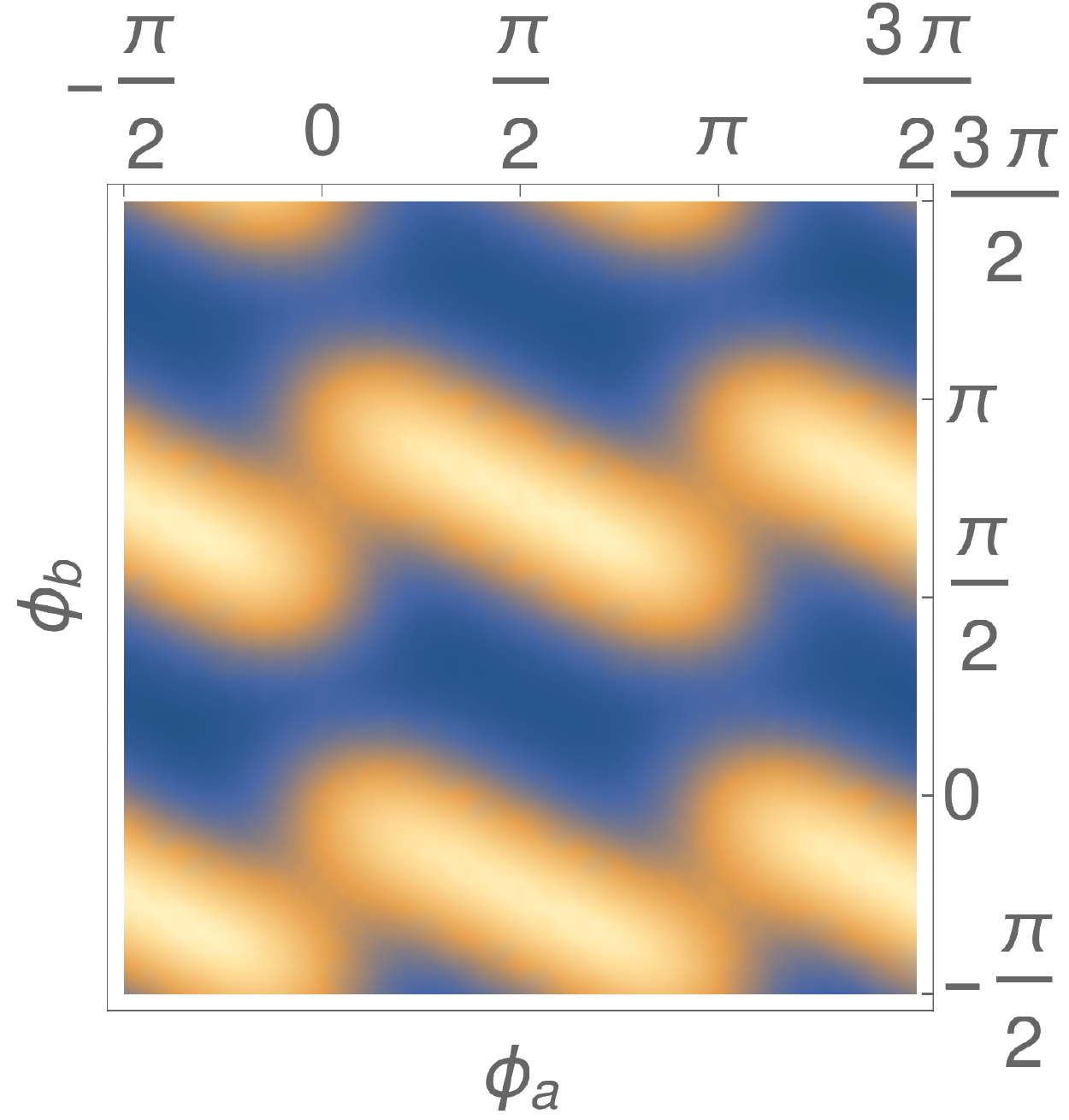}

}
\vspace{-.3cm}
\subfloat[Contribution to $Q'(\alpha,\beta)$ from trajectory in which transition
occured in $t_{1}=\pi/4$]{\includegraphics[width=0.15\textwidth]{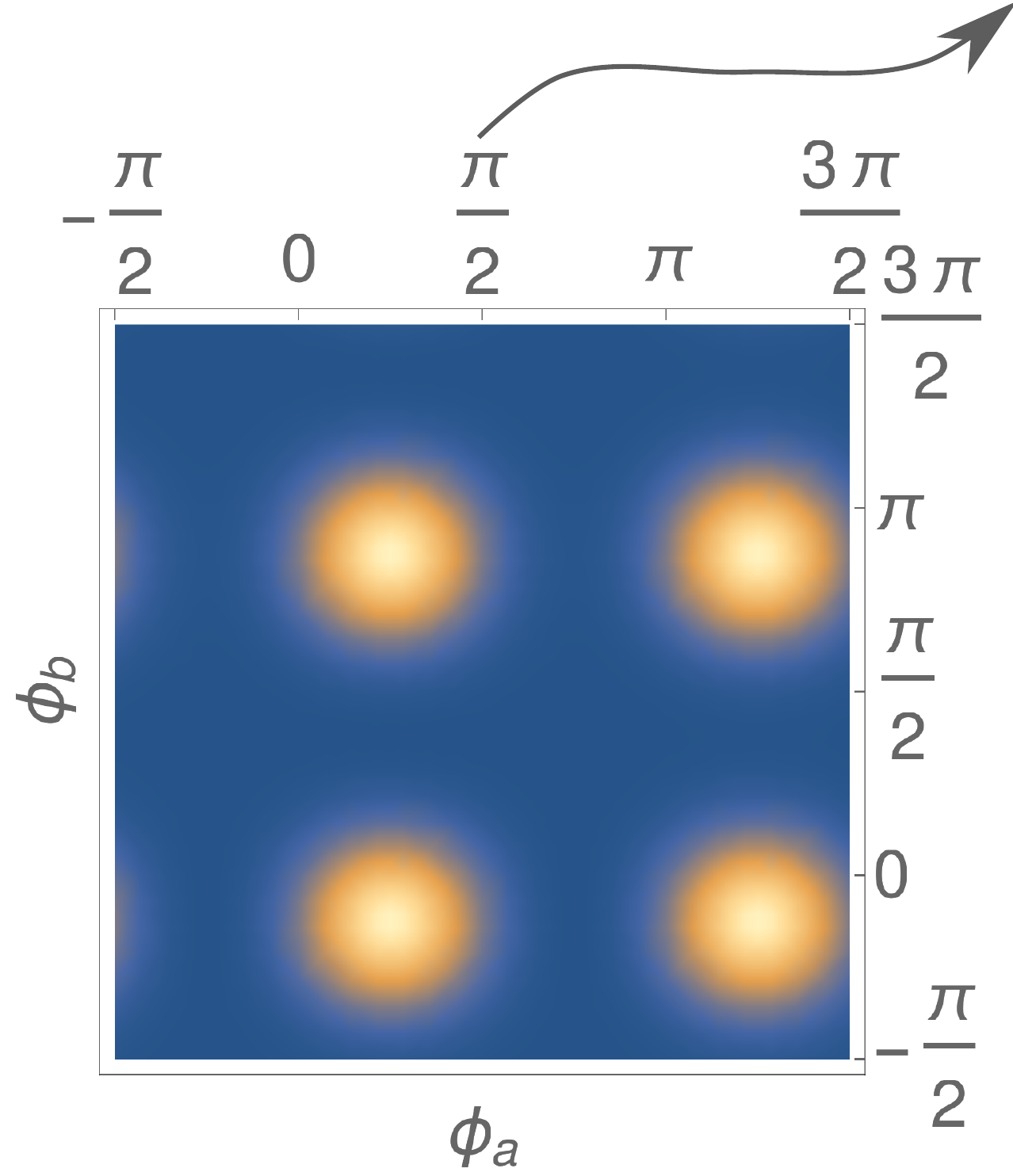}

}
\subfloat[Contribution to $Q'(\alpha,\beta)$ from trajectory in which transition
occured in $t_{1}=3\pi/4$]{\includegraphics[width=0.15\textwidth]{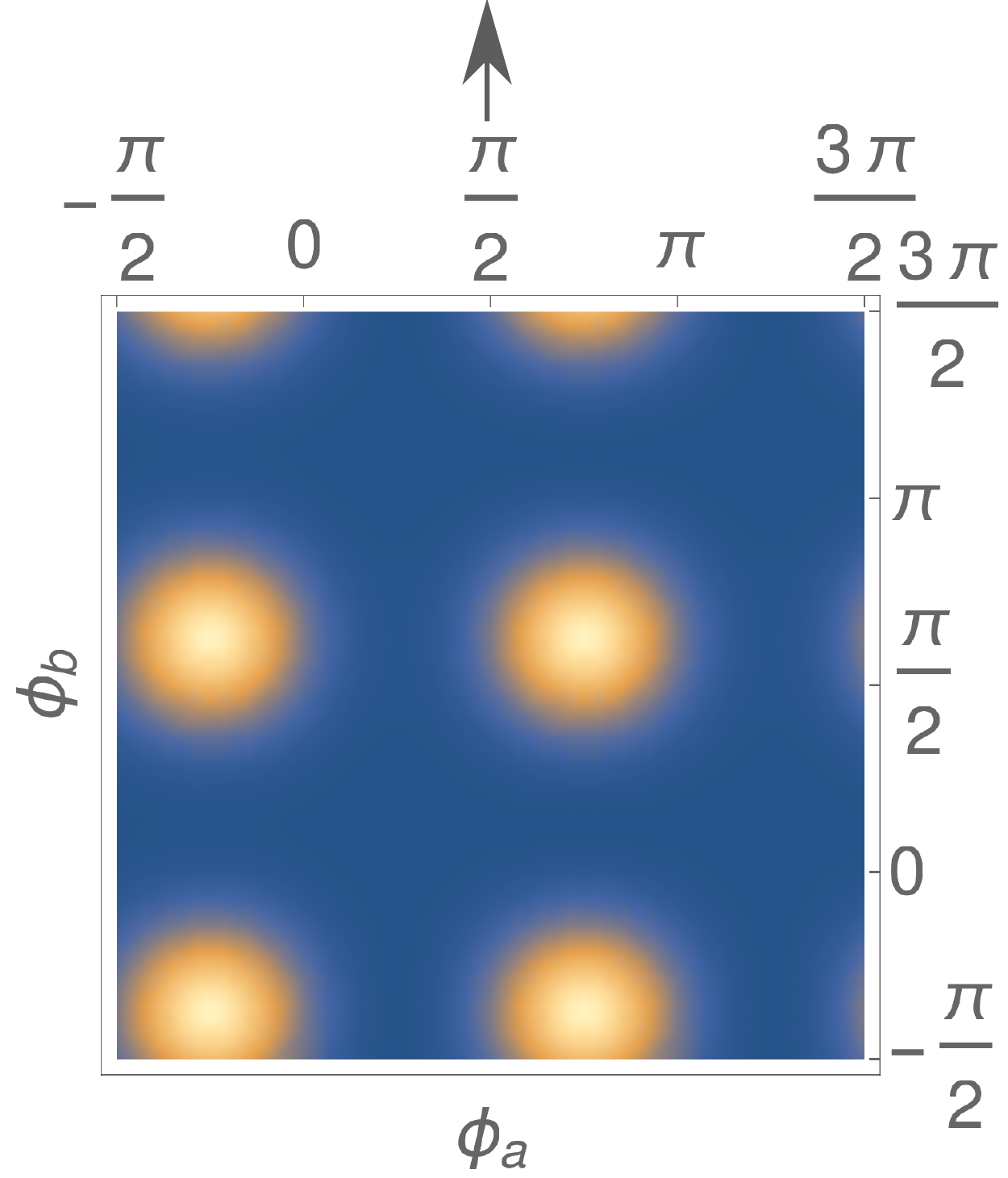}

}

\protect\protect\caption{Husimi Q functions for $N=10,\chi_{ab}=\chi$ , $\Gamma=0.01\chi$,
in the case where particles are lost only in the mode $\ket{0}$ of condensate
"a". The Husimi functions are computed in $\chi_{ab}t=\pi$.\label{fig:q}}
\end{figure}

In Fig. \ref{fig:q} (a) we present the cut of the Husimi function at time $\chi_{ab}t=\pi$ in the case without lost atom, where 
the state is the so called Schr\"odinger cat in two boxes, given already in Eq. \eqref{eq:sch-cat-in-two-boxes}.
In the panel (b) of Fig. \ref{fig:q} we visualize with the Husimi function the state \eqref{eqn:rhoprim}, restricted to the subspace with $N+M-1$ atoms. The state is smeared due to the phase noise described above.
Panels Fig. \ref{fig:q} (c) and (d) show the Husimi functions of single trajectories $\ket{\psi(t = \frac{\pi}{\chi_{ab}} |0, t_1)}$ for two chosen times $t_1$.

Due to the phase noise the Schr\"odinger cats disappear upon single atom  loss - the time of loss event is of the order of $t_1 \sim \pi/\chi_{ab}$ hence the resulting phase-noise is of the order of $\delta \phi \approx \chi_{ab}t_1 \approx \pi$. 
On the other hand, the EPR entanglement is relatively robust simply because it appears early enough, at time-scales $\tau_{\rm ent} \approx 1/ (\chi_{ab} N^{2/3})$. The average number of lost atoms during this time is around $N_{\rm loss}\approx \Gamma N \tau_{\rm ent}$, hence the number of lost atom scales like
$N_{\rm loss} \sim \Gamma N^{1/3}/\chi_{ab} $. Although the number of lost atoms growths when the total number of atoms is increased, but the total phase noise decreases:
$\Delta\phi \sim N_{\rm loss} (\chi_{ab} \tau_{\rm ent} ) \sim (\Gamma /\chi_{ab} ) N^{-1/3}$.

\section{Connection with Bose\textendash Einstein condensates \label{sec:bec}}

\subsection{Hamiltonian}
\begin{figure}
	\includegraphics[width=0.5\textwidth]{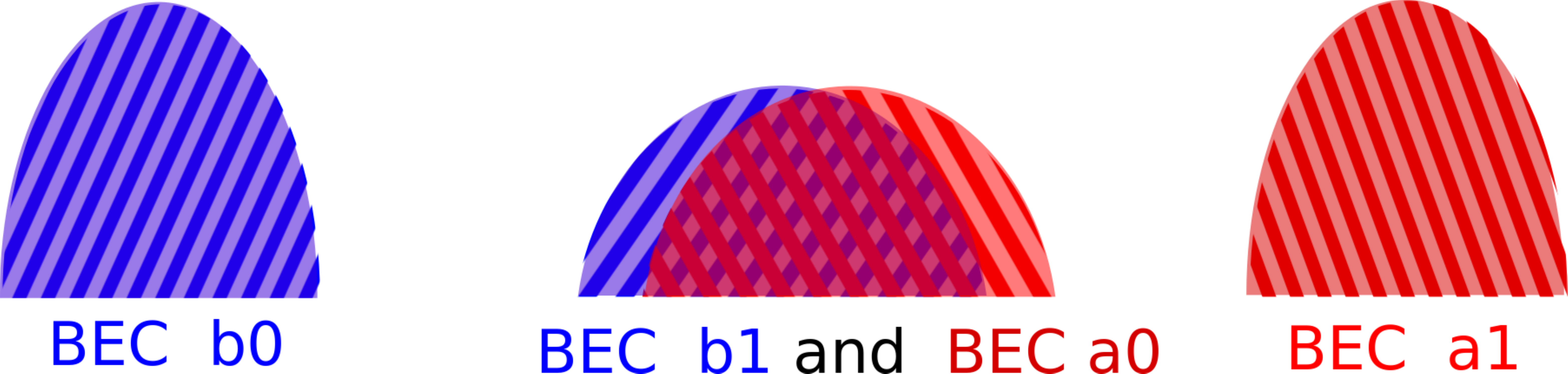}
	\caption{
		The scheme showing the central stage of the protocol to create
		the EPR-entangled states based on four Bose-Einstein condensates trapped
		in state-dependent potentials. As sketched in the plot, we assume that all
		BECs are in the Thomas-Fermi regime, hence they have parabolic shapes.
		The condensates "a0" and "b1" have lower densities than condensates "a1"
		and "b0" due to repulsive interactions between different species, i.e. collisions
		$\ket{0}$-$\ket{1}$. After the interaction time, the traps holding BECs should be moved
		to reach two bimodal condensates, "a" and "b". Only then the measurements
		should be taken to prove the EPR entanglement.
		\label{fig:schemeBEC4}}
\end{figure}

One of the scheme proposed as physical implementation of the Hamiltonian \eqref{eqn:ham0} is based on the clouds of atoms cooled down for the Bose-Einstein condensated to appear\cite{hadrien2013, hadrien2017}. 
The Bose-Einstein condensates are well isolated quantum systems with practically all parameters tunable.  Hence they are good candidates to test foundation of quantum mechanics and to implement the quantum information proposals.
Here we will briefly sketch the main results regarding coherent evolution.

The scheme \cite{hadrien2013} is based on two Bose-Einstein condensates, "a" and "b", each consisting of two-level atoms with internal states denoted with $\ket{0}, \ket{1}$. 
Each of the four components can be independently manipulated using appropriate state-dependent potentials \cite{Riedel2010}.
In the central stage of the scheme, atoms  in the states $\ket{0}$ from condensate "a" overlap with atoms in the state $\ket{1}$ from condensate "b" as shown in Fig. \ref{fig:schemeBEC4}.
To model mathematically the Hamiltonian of such four gaseous clouds, we will benefit from the simple form of the wavefunctions of BECs.
Namely, the many body wave-function $\psi_N \bb{\bm{r}_1, \bm{r}_2, \ldots, \bm{r}_N}$ of a condensate to a good approximation  is just equal to a single-body wave-function (orbital) $\psi^{({\rm GPE})}(\bm{r})$ occupied by all atoms, namely  $\psi_N \bb{\bm{r}_1, \bm{r}_2, \ldots, \bm{r}_N}= \prod_{i=1}^N \psi^{({\rm GPE})}(\bm{r}_i)$. In case of the four BECs, the system has to be described with four orbitals $\psi_{\sigma\epsilon}^{({\rm GPE})}$, where $\sigma = a,\,b$ and $\epsilon = 0,\,1$. 
In the stationary situation the orbitals can be calculated with the help of the four coupled Gross-Pitaevskii equations:
	\begin{equation}
	\mu_{\epsilon \sigma} \psi_{\epsilon \sigma}(\bm{r}) = \bb{\hat{T} +V_{\epsilon \sigma} +
		\sum_{\epsilon' \sigma'} g_{\epsilon' \sigma'} {N}_{\epsilon' \sigma'}  |\psi_{\epsilon' \sigma'}|^2 }\psi_{\epsilon \sigma}(\bm{r}) \label{eq:gpe4},
	\end{equation}
where $\hat{T} = -\frac{\hbar^2 }{2 m} \Delta$ is the kinetic energy operator, $V_{\epsilon \sigma} = \frac{1}{2} m \omega^2 \bb{x^2 + y^2 + (z-z_{\epsilon \sigma})^2}$ is the potential trapping the BEC "$\epsilon \sigma$" centered at $z_{\epsilon \sigma}$ and $g_{\epsilon \sigma}$ is the coupling constant for two-body collisions of atoms in internal states $\ket{\epsilon}$ $\ket{\epsilon'}$.
The energy of such four BECs is approximately
	\begin{align}
	E_{GPE}  &=   \sum_{\sigma\epsilon} N_{\epsilon \sigma} \int \psi_{\epsilon \sigma}(\bm{r}) \bb{\hat{T} + V_{\epsilon \sigma}}  \psi_{\epsilon \sigma}(\bm{r}) \nonumber\\
	&+ \frac{1}{2}\sum_{\epsilon \sigma\epsilon' \sigma'}N_{\epsilon \sigma}N_{\epsilon'\sigma'}|\psi_{\epsilon' \sigma'}(\bm{r})|^2 \, | \psi_{\epsilon \sigma}(\bm{r})|^2\label{eq:enGPE}.
	\end{align}
The equations \eqref{eq:gpe4} and \eqref{eq:enGPE} describe the BECs orbital and BECs energy in situation where the number of atoms in each from the four components is fixed , given by the integers $\{N_{\epsilon\sigma}\}$. 
The initial state considered in this paper is a superposition of Fock states with $N+M$ atoms differently distributed between the four modes. The full description of such system is quite involved \cite{hadrien2017}.  
Here we restrict the analysis to a simplified model, which stems from the formula for BECs energy \eqref{eq:enGPE} expanded in the Taylor series around the average values of the BECs occupation $\bar{N}_{\epsilon a} = N/2$ and   $\bar{N}_{\epsilon b} = M/2$:
	\begin{equation}
	\hat{H} \approx \chi_{\sigma} \bb{S_z^{\sigma}}^2 +  \chi_{ab} S_z^{a}S_z^{b}+ \sum_{\sigma} {\nu}_{\sigma} S_z^{\sigma} +\tilde{\chi}_{\sigma} N_{\sigma}S_z^{\sigma} + \bar{E}_{GPE}
	\label{eq:HAMapprox}
	\end{equation}
	The parameters $\chi_{\sigma}$, $\tilde{\chi}_{\sigma}$, $\chi_{ab}$ are combinations of the second derivatives of the energy $E_{GPE}$ with respect to the number of atoms:
	\begin{align}
		\chi_{\sigma} & =  \frac{1}{2\hbar}(\chi_{0\sigma} + \chi_{1\sigma} - 2 \chi_{01\sigma })\label{eq:chiGeneral}\\
		\chi_{0\sigma} & =  \frac{\partial^2 E_{GPE} }{\partial N_{0\sigma}^2} ;\quad \chi_{1\sigma }  =  \frac{\partial^2 E_{GPE} }{\partial N_{1\sigma}^2} ;\quad \chi_{01\sigma }  =  \frac{\partial^2 E_{GPE} }{\partial N_{0\sigma } \partial N_{1\sigma }} \nonumber\\
		\chi_{ab} & =  \frac{\partial^2 E_{GPE} }{\partial N_{0a} \partial N_{1b}} ; \quad 		\tilde{\chi}_{\sigma}  =  \frac{1}{2\hbar}(\chi_{1\sigma}-\chi_{0\sigma} ) \nonumber
	\end{align}
		
In the symmetric situation, when $N=M$ and the  coupling constants $g_{00}$ and $g_{11}$ are equal (which is close to the real situation in Rubidium-87), the Hamiltonian reduces to the form presented until now in our paper, Eq. \eqref{eqn:ham0}.
Comparison between the evolution given by the Hamiltonian \eqref{eq:HAMapprox} and the more involved model in which the spatial and time dependence of the orbitals $\psi^{({\rm GPE})}$ is accounted for is given in \cite{hadrien2017}.
	
To estimate analytically the Hamiltonian parameters for a real systems we follows the  papers
\cite{liyun2008, Li2009}.
In the limit of large number of atoms, the kinetic terms in the Gross-Pitaevskii equations \eqref{eq:gpe4}  can be neglected. This is the standard  Thomas-Fermi approximation. 
Then the equations \eqref{eq:gpe4} reduce to a set of algebraic equations, which in the case of symmetric couplings $g_{00}=g_{11} =: g$ can be solved analytically 
(due to  symmetry one has $\psi_{a0}(\bm{r}) = \psi_{b1}(\bm{r})$ and, up to a translation, $\psi_{a1}(\bm{r}) = \psi_{b0}(\bm{r})$). Having analytical form of the orbitals $\psi_{\epsilon\sigma}$ one can use the formula for GPE-energy \eqref{eq:enGPE} to derive the coefficients of the Hamiltonian
\eqref{eq:chiGeneral}. The results can be in fact deduced from the paper \cite{Li2009} (page 373):
\begin{align}
\label{eq:chi}
	\chi & = \frac{1}{5}\bb{\frac{15a}{2 l_{\rm osc}}}^{2/5} \bb{1+ \bb{\frac{a}{a + a_{01}}}^{3/5}} \omega \,N^{-3/5},\\
	\label{eq:chiab}
	\chi_{ab} & = \frac{2}{5}\bb{\frac{15a_{01}}{2 l_{\rm osc}}}^{2/5} \bb{\frac{a}{a + a_{01}}}^{3/5} \omega \,N^{-3/5} ,
\end{align}
where $a=\frac{m}{4\pi \hbar^2} g$ and $a_{01}=\frac{m}{4\pi \hbar^2} g_{01}$ are the scattering lengths, $m$ is the mass of a single atom, $\omega$ and $l_{\rm osc}=\sqrt{\hbar/(m\omega)}$ are the trap frequency and oscillatory length, respectively.

\subsection{Many-body losses}

\begin{figure}
	\includegraphics[width=0.4\textwidth]{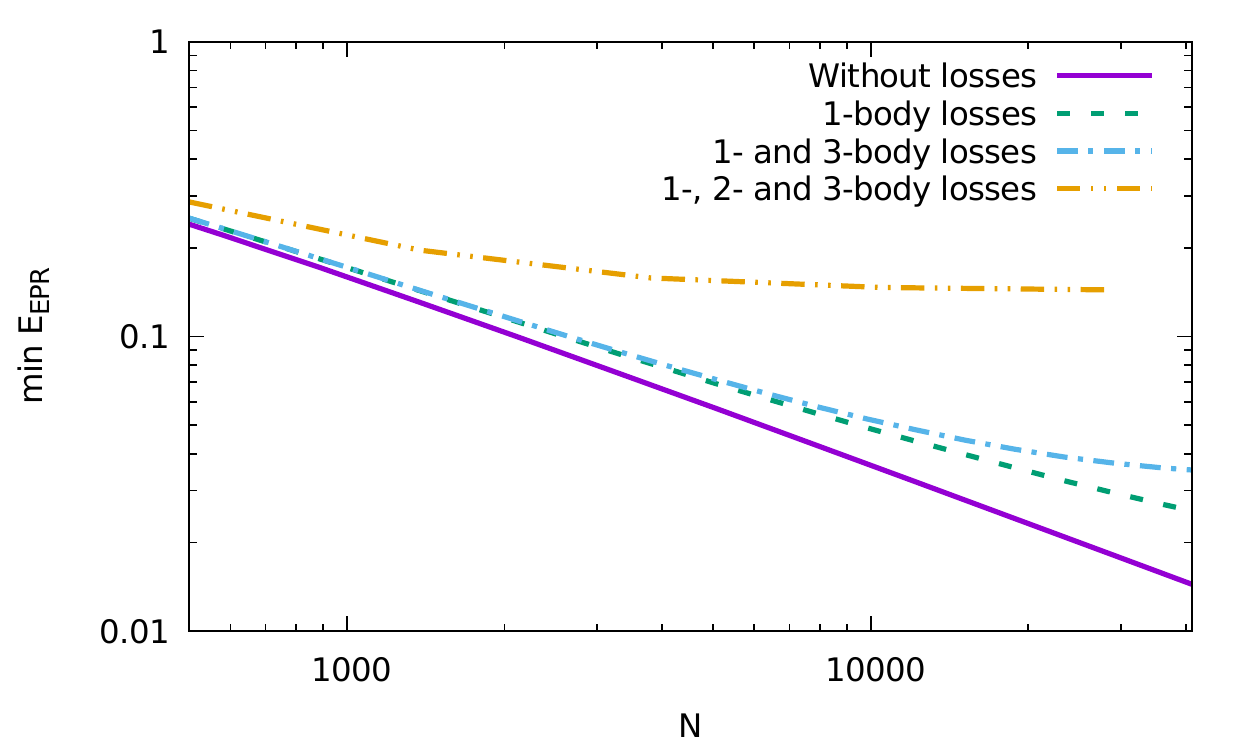}
	\includegraphics[width=0.4\textwidth]{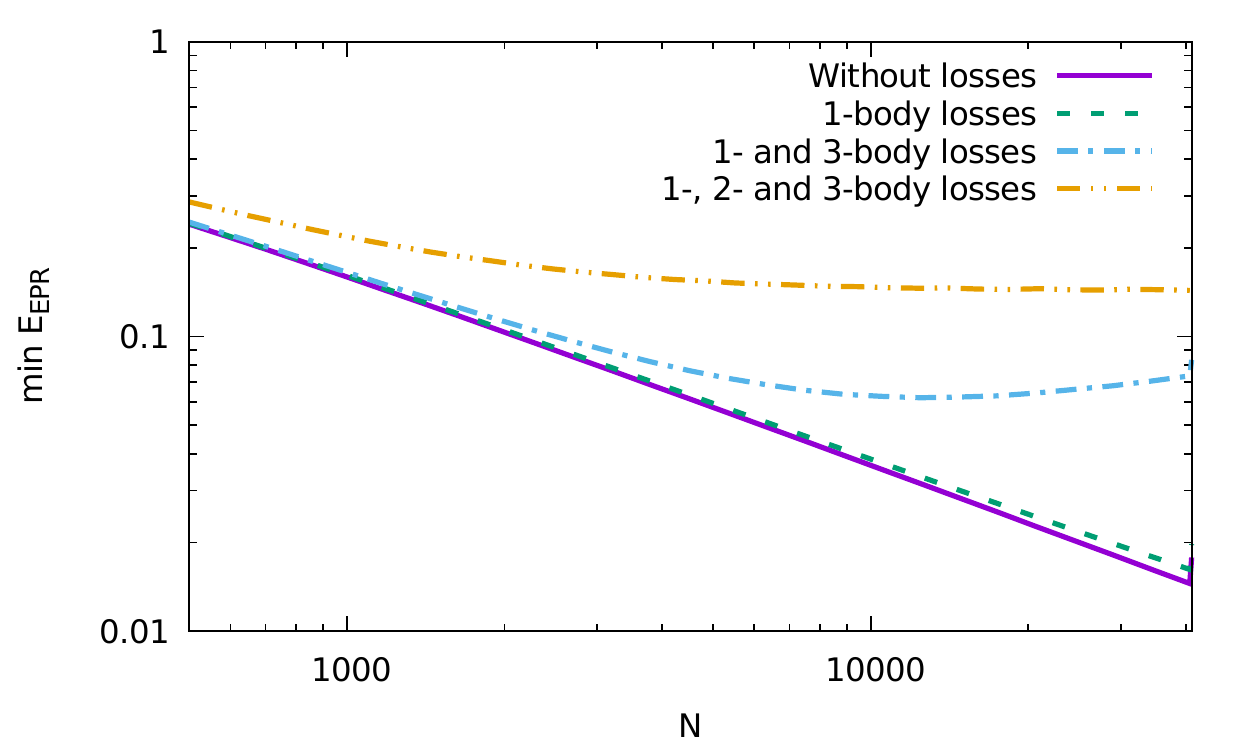}
	\caption{Plot of the minimum $E^2_{EPR}$ as a function of total number of particles. The atoms are assumed to be symmetrically distributed among the two harmonic traps with the trap frequency: {\bf top} $\omega = 2\pi\times 200$Hz and {\bf bottom} $\omega = 2\pi\times 1000$Hz. 
		The scattering length are equal to $a_{01} = a = 100$ Bohr radii and mass is $m=87$ in atomic units.
		All two- and three- body losses constants were assumed to be the same and  equal to $K_{01}^{(2)} = 8\times 10^{-20}$m$^3$/s and $K^{(3)}=6\times10^{-42}$ m$^6$/s ,respectively.
		The rate of one body losses was assumed to have the (non-challenging) value $K^{(1)}=0.5$Hz.
		\label{fig:minen}}
\end{figure}

Real life condensates do note evolve in accordance to Hamiltonian only: losses are present and often are a limiting factor in experiments. From the experimental standpoint, most important types of losses are
\begin{enumerate}
\item one-body losses, arising from interaction with ambient molecules (imperfect vacuum) -- the jump operators are linear in ladder operators $\hat{a}_0, \hat{a}_1, \hat{b}_0, \hat{b}_1$, as described before
\item two-body losses, when the interaction between two atoms makes them change spin  -- the jump operators are quadratic. For $^{87}$Rb  spin states $\ket{F = 1, m_F = -1}=\ket{0},\ket{F = 2, m_F = 1}=\ket{1}$ the possible operators are $A^{(2)}_{11} =\sqrt{\gamma_{11}^{(2)}}\,\hat{a}_1 \hat{a}_1$, $A^{(2)}_{01} = \sqrt{\gamma_{01}^{(2)}}\hat{a}_0 \hat{a}_1$ and analogously for "b" subsystem.
The conservation of the magnetization during spin exchange collision prevent  $0$-$0$ losses, i.e. $\gamma_{00}^{(2)} = 0$.
\item three--body losses, happening during inelastic three body collision during which one of the interacting atoms gain enough energy to leave the condensate and two others form a stable bimolecule. 
The jump operators corresponding to these losses are given by the third order polynomials of annihilation operators, for instance all losses in "a" are  of the form $A^{(3)}_{\epsilon_1\epsilon_2\epsilon_3}=\sqrt{\gamma_{\epsilon_1\epsilon_2\epsilon_3}^{(3)}} \hat{a}_{\epsilon_1}\hat{a}_{\epsilon_2}\hat{a}_{\epsilon_3}$.
\end{enumerate}
The two- and three-body losses are originated in the two- and three-body collisions, respectively, hence they depend on the atomic densities. For instance the rates $\gamma^{(m)}$ of $m$-body losses in the component "a" are given by
	\begin{equation}
	\gamma_{\epsilon_1 \ldots \epsilon_m}^{(m)} = 
	\frac{K_{\epsilon_1 \ldots \epsilon_m}^{(m)}  }{m} \int  |\phi_{\epsilon_1,a}|^{2}\ldots  |\phi_{\epsilon_m,a}|^{2}  {\rm d}^3 r,
	\end{equation}
where $K_{\epsilon_1 \ldots \epsilon_m}^{(m)} $ are atomic constants.

The evolution in the presence of one-body losses is given by the master equation \eqref{eq:master}, but with
$\chi$ and $\chi_{ab}$ being functions of the total number of atoms, see equations \eqref{eq:chi}, \eqref{eq:chiab}.
From the preceding analysis of time-scales we expect that the number of atoms lost until the $E_{\rm EPR}$ reaches minimum will be small.
Hence we neglect the time dependence of $\chi$ and $\chi_{ab}$ resulting from the changes in time of the total number of atoms $N$ and $M$. 
In this model, in which  $\chi$ and $\chi_{ab}$ are function of the total initial number of atoms, the conclusions from the previous parts do not holds.

Taking into considerations the effect of number of particles on parameters, we may plot the minimum $E^2_{EPR}$ as a function of $N$, which we take for simplicity to be equal to $M$, see Fig. \ref{fig:minen}.
Each point from this figure has been obtained by  optimizing with respect to $\alpha$, $\beta$  and time in formula \eqref{eq:def-eepr}. The atomic constants used in these plots are close to the parameters of Rubidium 87. In the upper panel of Fig. \ref{fig:minen} , corresponding to trap frequency $\omega=2\pi\times 200$Hz, one can see  that for increasing number of atoms the effect of 1-body losses (green dashed line) is stronger and stronger. This comes from the fact that the nonlinearities $\chi$ and $\chi_{ab}$ tend to $0$ with the number of atoms. The time to reach the minimum of $E_{\rm EPR}$ increases what, for fixed  1-body loss rate constant, lead to severe losses in the limit $N\to \infty$. 
To counteract this effect one should increase the gas density to enhance the nonlinearities and make the unitary evolution faster, as illustrated in the lower panel of Fig. \ref{fig:minen} , corresponding to trap frequency $\omega=2\pi\times 1000$Hz.
The side effect of making the confinement steeper are increased density-dependent losses, i.e.  two and three-body losses.

We covered the existence of one-body losses in fully analytical way. 
However, the method of characteristic functions does not work  even for two-body losses as it leads to the partial differential equations of order 2 with unknown solutions. Therefore, a procedure for heurestic incorporation of higher-order losses into one body losses was devised. The prescription is deduced from the method of quantum trajectories:
we increase the rate of one--body losses such that the average number of lost atom at the interval $t, t+\Delta t$ is the same as there would be all losses present. 
Precisely the number of atoms lost in BEC "a" with the internal state $\ket{0}$ in the small time interval $\Delta t$ is equal to  
	\begin{equation}
	N_{a0}(t+\Delta t) - N_{a0}(t) \approx \sum_{m=1}^3\sum_{\bm{\epsilon}} n_{a0, \bm{\epsilon}} \left\langle \bb{\hat{A}_{\bm{\epsilon}}^{(m)}}^{\dagger}\hat{A}_{\bm{\epsilon}}^{(m)}\right\rangle \Delta t,
	\end{equation}
where $\hat{A}_{\bm{\epsilon}}^{(m)}$ are jump operators corresponding to $m$-body losses, and
$n_{a0,\bm{\epsilon}}$ is the number of lost atoms in the component "a0" due to the jump $\hat{A}_{\bm{\epsilon}}^{(m)}$.
We approximate the averages of $m$-body operators $\left\langle \bb{\hat{A}_{\bm{\epsilon}}^{(m)}}^{\dagger}\hat{A}_{\bm{\epsilon}}^{(m)}\right\rangle $ with 
an average of $1$-body operator $\Gamma_0^{\rm eff}\langle \hat{a}^{\dagger}_{0}\hat{a}_{0}\rangle$ with the effective rate $\Gamma_0^{\rm eff}$ chosen in such a way that at least in the limit of weak losses and short evolution times, when $\langle N_{a1}\rangle \approx \langle N_{a0}\rangle\approx N/2$,  the total rate of loosing particles will be similar in both models \footnote{Technically, we contract operators. For example 
	$\langle \Gamma_{001}^{(3)} \hat{a}^{\dagger}_{0}\hat{a}^{\dagger}_{0}\hat{a}^{\dagger}_{1}\hat{a}_{0}\hat{a}_{0}\hat{a}_{1} \rangle \mapsto 
	\Gamma_{001}^{(3)} \bb{\frac{N}{2}}^2\langle  \hat{a}^{\dagger}_{0}\hat{a}_{0}\rangle $.}.
This leads to the following formula for the effective rate:
	\begin{equation}
	\Gamma_{0}^{\rm eff}  =  \gamma_{\epsilon} + N \, \bb{\gamma_{11}^{(2)} + \frac{1}{2} \gamma_{01}^{(2)} } + \frac{3}{4}\gamma^{(3)} N^2.
	\end{equation}
	
As the loss rates constants $K_2$ are asymmetric \cite{pawlowski2017} and , as shown schematically in Fig. \ref{fig:schemeBEC4}, the densities of overlapping BECs are different from the BECs in the outer traps, then
 each from the four BECs would have a different rate of losses. We do not account this asymmetry.
Keeping in mind that our analysis of the many-body losses is only qualitative, we will do the analysis for symmetric rates, approximating their values within Thomas-Fermi approximation \cite{Li2009}:
\begin{align}
\gamma^{(2)} &= \frac{1}{14 \pi}\bb{\frac{15}{2}}^{2/5} \frac{K_2}{l_{\rm osc}^3} N^{-3/5} \bb{\frac{l_{\rm osc}}{a}}^{3/5}\\
\gamma^{(3)} &= \frac{1}{126 \pi^2}\bb{\frac{15}{2}}^{4/5} \frac{K_3}{l_{\rm osc}^6} N^{-6/5} \bb{\frac{l_{\rm osc}}{a}}^{6/5}
\end{align}
To keep things simple and use symmetry we assume in the formulas above that all BECs have the density like "a1" and "b0" clouds, and that the two body losses are present in all components with the rate constant $K_2$. In this way we enhance the losses, so the following results are the black scenario. 

In Fig. \ref{fig:minen} we show the results which include phenomenologically the two and three body losses according to the prescription described above.
As opposed to the result without proper scaling, there exists an optimal number of particles for which the minimum $E^2_{EPR}$ is attained. It is easily explained heurestically: in the limit of large number of atoms, the density has to increase. In high densities regime however the atoms scatter very often, what leads to increased number of three-body losses. One could counteract this effect by decreasing the trap frequency, this leads to small coefficients $\chi,\chi_{ab}$ though and EPR entanglement occurs at later times, in this case the system is limited by one-body losses. The mechanism is similar to the limits on the squeezing in Bose-Einstein condensates discussed in \cite{liyun2008}.

Summarizing this part: in our heuristic analysis including $2$ and $3$-body losses, where we assumed stronger losses than should be in reality, still one can obtain EPR-entangled condensates consisting of thousands of atoms, as shown in Fig. \ref{fig:minen}.


\section{Conclusions}
We presented the analytical solution of the master equation describing the two macroscopic spins interacting via $\sza\szb$ term, but undergoing the one-body particle losses. 
The intuitive picture is gained via the quantum trajectory method, which shows that the mechanism underlying the decoherence is the phase noise, which here is non-local.
We discussed the fate of the non-local entangled states predicted in the unitary evolution. As expected, the so called "Schr\"odinger cat in two boxes" loose the quantum coherence once a single particle is lost.
Much more robust is the entanglement captured by the EPR condition, appearing at short evolution times. We discuss possibility of producing them from the perspective of the Bose-Einstein condensates.  

We hope that our work can be used in the Quantum Information community as a simple, still rich, model with the analytical solutions even in the presence of particle losses. It is known that the $\sza\szb$-scheme can lead to many interesting , potentially useful, entangled states \cite{byrnes2013devil}.
The next steps in research could be finding an optimal situations for production of the EPR state in BEC taking into account all losses (we touched the problem  in Sec. \ref{sec:bec}) or trying to use another entanglement witness to find another  correlated states appearing in the dissipative evolution.

\acknowledgments
This work was supported by the (Polish) National Science Center Grants 2014/13/D/ST2/01883 (K.P. and K.S.).

\appendix
\section{Characteristic function\label{app:characteristic-function}}

\begin{widetext}
The master equation \eqref{eq:master} written in the Fock basis, namely $\frac{d}{dt}\langle x+z,y,u+r,v | \hat{\rho}| x,y+z,u,v+r\rangle $, read
\begin{equation}
\begin{split}
\frac{d}{dt}\rho_{x+z,y,u+r,v}^{x,y+z,u,v+r}= & i\left(\chi_{a}4z(y-x)+\chi_{b}4r(v-u)+2\chi_{ab}\left(r(y-x)+z(v-u)\right)\right)\rho_{x+z,y,u+r,v}^{x,y+z,u,v+r}-\\
  &~~\left(\Gamma_{0}(2x+2u+z+r)+\Gamma_{1}(2y+2v+z+r)\right)\rho_{x+z,y,u+r,v}^{x,y+z,u,v+r}+\\
  &\quad\Gamma_{0}\left(\rho_{x+z+1,y,u+r,v}^{x+1,y+z,u,v+r}\sqrt{(x+1)(x+z+1)}+\rho_{x+z,y,u+r+1,v}^{x,y+z,u+1,v+r}\sqrt{(u+1)(u+r+1)}\right)+\\
  &\quad\Gamma_{1}\left(\rho_{x+z,y+1,u+r,v}^{x,y+z+1,u,v+r}\sqrt{(y+1)(y+z+1)}+\rho_{x+z,y,u+r,v+1}^{x,y+z,u,v+r+1}\sqrt{(v+1)(v+r+1)}\right).
\end{split}
\end{equation}
Hence for fixed $z$ and $r$ we obtain a set of a coupled differential equation on the density matrix terms $\rho_{x+z,y,u+r,v}^{x,y+z,u,v+r}$ labeled with $x$, $y$, $u$ and $v$. This set can be elegantly solved with help of the characteristic function:
\[
h^{z,r}(X,Y,U,V,t)=\sum_{x,y,u,v}\sqrt{\frac{(x+z)!(y+z)!(u+r)!(v+r)!}{x!y!u!v!}}X^{x}Y^{y}U^{u}V^{v}\rho_{x+z,y,u+r,v}^{x,y+z,u,v+r}.
\]
By summing the equations for individual $\rho_{x+z,y,u+r,v}^{x,y+z,u,v+r}$ with weights $\sqrt{\frac{(x+z)!(y+z)!(u+r)!(v+r)!}{x!y!u!v!}}$ one obtain a closed, equation on the characteristic function \eqref {eq:characteristic-function}:
\begin{align}
\frac{\partial h^{z,r}}{\partial t}= & \left(-\beta^{(r,z)}_0X+\Gamma_{0}\right)\frac{\partial h^{z,r}}{\partial X}+
\left(\beta^{(r,z)}_1 Y+\Gamma_{1}\right)\frac{\partial h^{z,r}}{\partial Y}+ \nonumber\\
&  \left(-\beta^{(z,r)}_0U+\Gamma_{0}\right)\frac{\partial h^{z,r}}{\partial U}+
\left(\beta^{(z,r)}_1 V+\Gamma_{1}\right)\frac{\partial h^{z,r}}{\partial V} \nonumber\\
& -  2\left(\Gamma_{0}+\Gamma_{1}\right)(z+r)h^{z,r}.
\end{align}
This first order partial differential equation can be solved with the standard methods, as the  method of characteristics.
For the initial state investigated in the paper the final result is
\begin{equation}
h^{z, r} (X, Y, U, V, t) = \frac{N! M! e^{-\frac{1}{2} \left(\Gamma_0+\Gamma_1\right)(r+z)\, t }}{ 2^{r+z}(M-r)! (N-z)!} \, \bb{L_{z,r}(X,Y,t)}^{N-z}\, \bb{ L_{r,z} (U,V,t)}^{M-r},
\label{eq:hzr-explicite}
\end{equation}
where
\begin{eqnarray}
L_{z,r}(X,Y,t) &=& \frac{1}{2}\bb{\frac{\Gamma_0 + \bb{A_{z,r} X - \Gamma_0}e^{-A_{z,r} t}}{A_{z,r}} +\frac{\Gamma_1 + \bb{B_{z, r} Y - \Gamma_1}e^{-B_{z, r} t}}{B_{z, r}}}\\
A_{z, r} &=& \Gamma_0 + i z \chi + i r \chi_{ab}/2\\
B_{z, r} &=& \Gamma_1 - i z \chi - i r \chi_{ab}/2
\end{eqnarray}

\section{Quantum averages\label{app:quantum-averages}}

Using the characteristic function \eqref{eq:hzr-explicite}, all number-of-particles-preserving correlators can be calculated (the others are $0$ in our system). 
For example, $\langle a_1^\dagger a_0 \rangle = h^{1,0}\mid_{X=Y=V=1}$.
Unfortunately, the explicit equations are complicated and would not fit on one page, so we will define the averages by intermediate functions. Therefore, we define
\begin{equation}
\alpha^{(z,r)} = z\, \chi + r \chi_{ab} / 2
\end{equation}
and subsequently
\begin{equation}
f_{z,r}(t)=L_{z, r}\mid_{X=Y=U=V=1} =\frac{1}{2}\left(\frac{ \Gamma_0 + i \alpha^{(z,r)} e^{ -\Gamma_0 t - i t\alpha^{(z,r)} }  }{\Gamma_0 + i \alpha^{(z,r)}}+\frac{\Gamma_1 - i \alpha^{(z,r)} e^{-\Gamma_1 t + i t \alpha^{(z,r)}}}{\Gamma_1 - i \alpha^{(z,r)}}\right)
\end{equation}
The quantum averages necessary to evaluate the steering condition $E_{\rm EPR}$ read:
\begin{align*}
\langle S^a_x \rangle &= \Re h^{1,0} \quad\quad \langle S^a_y \rangle = \Im h^{1,0} \quad\quad \langle S^a_z \rangle = \frac{N}{4}\bb{ e^{-t\Gamma_1}- e^{-t\Gamma_0 } }\\
 N(t) &= \frac{1}{2} N \bb{e^{-\Gamma_0 t} + e^{-\Gamma_1 t}}\\
 \langle \left(S^a_y\right)^2 \rangle &= \frac{1}{4} \bb{ N(t) + \frac{N(N-1)}{2} e^{-(\Gamma_0+\Gamma_1) t}\bb{1 - \Re \left\{f_{2,0}^{N-2}f_{0,2}^{M} \right\}}}\\
 \langle \left(S^a_z\right)^2 \rangle &=\frac{N (N-1)}{16}  \bb{e^{-\Gamma_1 t} - e^{-\Gamma_0 t}}^2  + \frac{1}{4} N(t)\\
 \langle S^a_y S^b_y \rangle &= -\frac{M N e^{-\left(\Gamma_0+\Gamma_1\right) t} }{8} \Re\bb{ f_{1,1}^{N+M-2} - f_{1,-1}^{N-1}f_{-1,1}^{M-1} } \\
\langle S^a_y S^b_z \rangle &= \frac{M N  e^{-\left(\Gamma_0 + \Gamma_1\right) t/2} }{8}
\Im \left\{  f_{1,0}^{N-1}f_{0,1}^{M-1}  \bb{ e^{ -\left(\Gamma_0 + i \chi_{ab}/2 \right) t} - e^{ -\left(\Gamma_1 -  i \chi_{ab}/2 \right) t} }  \right\} \\
 \langle \{ S^a_y, S^a_z \} \rangle &= \frac{N (N-1) }{4} e^{-\frac{1}{2}\left(\Gamma_0+\Gamma_1\right) t} \Im \left\{  f_{1,0}^{N-2} f_{0,1}^{M}
 \left( e^{-(\Gamma_1-i\chi) t}  -e^{-(\Gamma_0+i\chi) t} \right) \right\},
\end{align*}
\end{widetext}

\bibliographystyle{plain}
\bibliography{refs}
\end{document}